\documentclass{article}
\usepackage{amsmath}
\usepackage{graphicx}
\usepackage[numbers,sort&compress]{natbib}
\usepackage{cleveref}
\usepackage{booktabs}
\usepackage[table]{xcolor}
\definecolor{rowgray}{gray}{0.9}
\usepackage{siunitx}
\usepackage{url}

\usepackage{float}
\usepackage{tabularx}   
\usepackage{threeparttablex}
\newcolumntype{L}{>{\raggedright\arraybackslash}X}  

\title{Finding Inter-species Associations on Large Citizen Science Datasets}
\author{Jacob Deutsch\\
  \small Terrabyte Lab, Santa Cruz, California, USA\\
  \text{jacobadeutschwork@gmail.com}}
\date{August 2025}

\begin{document}
\maketitle


\begin{abstract}
\noindent
Determining associations among different species from citizen science dat- abases is challenging due to observer behavior and intrinsic density variations that give rise to correlations that do not imply species associations. This paper introduces a method that can efficiently analyze large datasets to extract likely species associations. It tiles space into small blocks chosen to be of the accuracy of the data coordinates, and reduces observations to presence/absence per tile, in order to compute pairwise overlaps.  It compares these overlaps with a spatial Poisson process that serves as a null model. For each species $i$, an expected overlap $\mu_i$ is estimated by averaging normalized overlaps over other species in the same vicinity.  This gives a $z$-score for significance of a species-species association and a correlation index for the strength of this association. This was tested on $874,263$ iNaturalist observations spanning $15,975$ non-avian taxa in the Santa Cruz, California region ($\approx 4.68\times10^{6}$ tiles). The method recovers well-known insect host-plant obligate relationships, particularly many host-gall relationships, as well as the relationship between Yerba Santa Beetles and California Yerba Santa. This approach efficiently finds associations on $\sim10^{8}$ species pairs on modest hardware, filtering correlations arising from heterogeneous spatial prevalence and user artifacts. It produces a ranked shortlist of ecological interactions that can be further pursued. Extensions to this method are possible, such as investigating the effects of time and elevation. It could also be useful in the determination of microhabitats and biomes.
\end{abstract}


\section{Introduction} 

Mapping inter-species interactions is fundamental to understanding how ecosystems function \cite{BascompteJordano2014}. Interactions have usually been studied one species pair at a time. The aim of this paper is to demonstrate a new method for uncovering inter-species interactions with large data sets. For example, iNaturalist's dataset consists of more than 200 million observations \cite{MasonEtAl2025_iNaturalistAccelerates}. Although some of these data points are annotated with respect to species interactions, the majority of iNaturalist data only contain three columns of relevant information. A taxon name, coordinate, and time. The overwhelmingly larger amount of this simpler information and what can be attained from it is the main subject of this research.

Many ecological‑niche modeling (ENM) techniques have been developed that use co-occurrence and environmental factors to make predictions \cite{barbosa2012mediterranean, thuiller2024niche, ElithLeathwick2009, Phillips2006Maxent}. An earlier approach, Deep Multi-Species Embedding (DMSE), used neural-network models on eBird datasets to predict inter-species co-occurrence \cite{chen2016deep}.  In addition, the analysis of photographs from citizen science projects have been used to extract species interactions \cite{groom2021species}. Notably many pollinator plant interactions have been revealed this way \cite{bosenbecker2023contrasting, gazdic2019inaturalist}. Studies continue to fill in ecological knowledge gaps using citizen science data from websites like iNaturalist \cite{putman2021power, hu2025global}. Despite the insight that has been gained by ecological-niche modeling, even the best ENMs can misestimate interaction potential between species \cite{wen2024invasive}. One reason for this could be the tendency to make lots of assumptions within their models. With such a complex system, minimizing the number of assumptions made should help improve accuracy.

Citizen science data is usually far less reliable than data taken in the course of scientific studies and it also has a significant user bias \cite{BrownWilliams2019, Isaac2014}. People tend to stay on trails, upload more iconic and obvious species, and mainly observe during daylight hours. Despite the lower quality of individual observations, there are many orders of magnitude more of them. The aim of this work is to come up with statistical techniques to handle data of this kind in order to extract useful biological information.

\section{Methods} 
The first and simplest way to quantify inter-species correlations is to check if two species' data points tend to be closer together than expected by chance. If that is the case, this is an indication that there may be some correlation between them, which may suggest a potential for an inter-species interaction.

We start by considering two species $i$ and $j$. Then we take an individual data point from species $i$ and draw a circle around it. We ask if the number of data points from species $j$ in that same circle is to be expected assuming that there is no interaction between species $i$ and $j$. I discuss this approach in the results section. However, it is much less computationally efficient than simpler methods. The primary goal of this research is to extract information about species associations from very large datasets, computing correlations for millions of species-species pairs over large regions. I have focused on a more efficient method that gives us very similar information by tiling the entire area containing the dataset and looking at the presence or absence of each species in each tile.

\begin{figure}[tb]
	\centering
	\includegraphics[width=\linewidth]{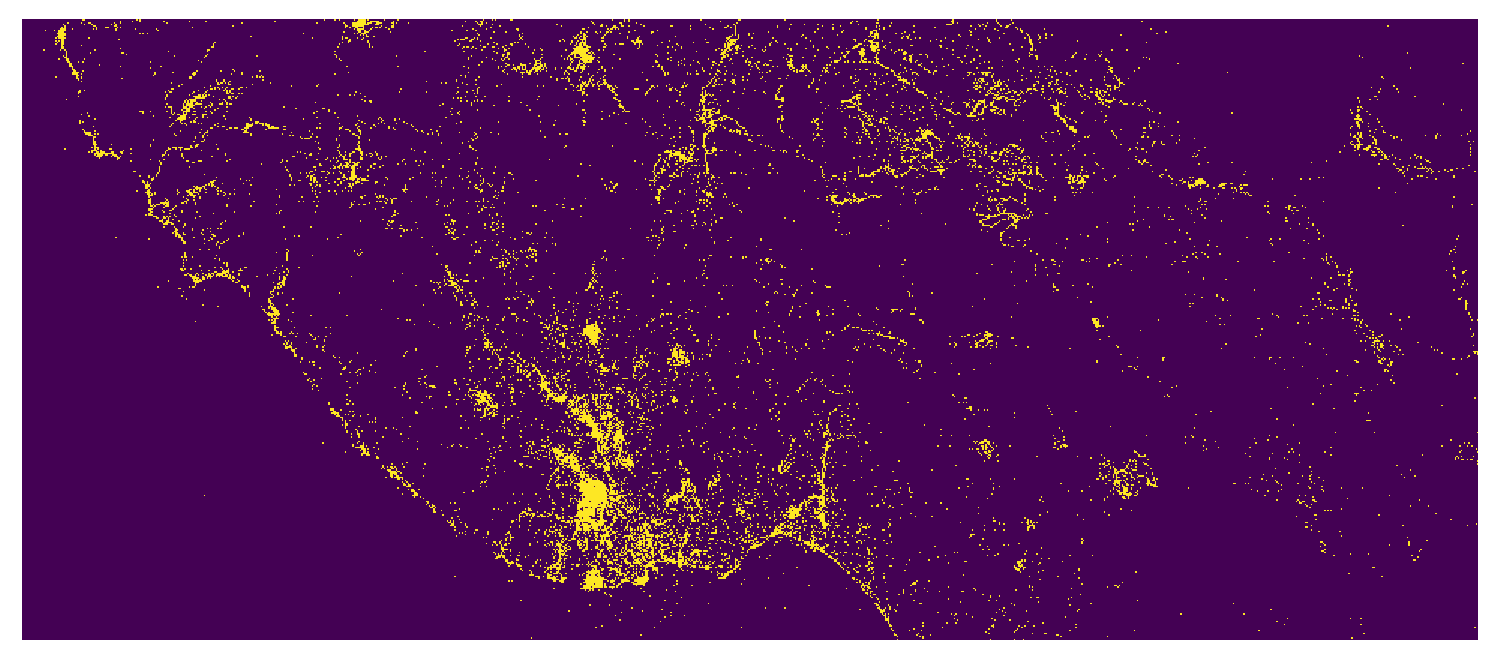}
	\caption{Tiled map of the Santa Cruz Mountains. The yellow tiles contain at least one observation. The purple tiles are empty.}
	\label{fig:map_yellow_purple_SC}
\end{figure}

The data itself is restricted to where users typically visit (see~Fig.~\ref{fig:map_yellow_purple_SC}). The clumping that is seen in the data is predominantly an artifact of the observers, and not an intrinsic ecological property. Take a wetland trail for example: Right in the same spot people might upload birds seen hundreds of meters away and might also upload small beetles crossing the trail. This is because the location data of the photos is predominantly confined to the trail. Most people will not try to fine tune the locations of all the birds they spot. This could easily lead to a seemingly significant but spurious relation between herons and marsh beetles, for example. An apparent correlation due to this sort of observational bias carries no useful ecological information. Of course they both inhabit the same ecosystem, but the goal here is to gain deeper insight into inter-species interactions. The primary focus of this method is determining a baseline to check whether the population of species $j$ in the vicinity of species $i$ is to be expected assuming there is no interaction between the two species.

The first step of this method is to tile the data into square areas. For this paper I chose tile sizes of \SI{33}{m} as an estimate of average location data accuracy \cite{ZandbergenBarbeau2011, MerryBettinger2019}.

Having multiple data points for the same animal leads to problems with the analysis: spuriously high correlations, and makes the data non-Poissonian. Only using the presence or absence of a species within a tile mitigates both of these issues. Therefore we employ a tiling method in which we mark present or absent per tile for each species rather than working with the raw number of data points.

The inter-species comparisons are then computed by going through each species pair in the dataset (if two species share at least one tile then they are compared). We define the overlap, which we denote as $O_{ij}$, as the number of tiles in which the two species both are present together.

Our null hypothesis is that the data for a single species is drawn from a spatial Poisson process \cite{baddeleybaranyschneider2007}. This would mean that correlations in the data that we see are due to spatial heterogeneity in observations and random noise, rather than a real association. More specifically, we consider two species $i$ and $j$. Given the data points for $i$, we ask if the data points for $j$ are well described as a spatial Poisson process. In other words, the quantities derived from the data, such as the overlap, $O_{ij}$, between $i$ and $j$, are what are expected typically for species in geographical proximity.

In order to determine if species $j$ has an atypically large number of tile overlaps with species $i$, one needs to first determine the typical overlap number. We can get a typical overlap number for a species $i$, which we will denote as $\mu_i$, by considering other species that share geographical proximity with species $i$.

To get an estimate of a typical overlap with a given species $i$, we can make use of the fact that there are almost always many other species that overlap with it. We can use these other species to get an estimate for a typical overlap. To be more specific, we are considering all tiles where species $i$ is present. For each of these we list the other species in the same tiles. Then we will obtain an average using these other species.

There is a problem caused by the varying tile count of species in the dataset. The total number of tiles occupied by a species varies greatly \cite{Baldridge2016_SADComparisonPeerJ}. This leads to artifacts when performing averages that are discussed further in section~\ref{sec:exploring_mui}. To  account for this, we normalize by the tile count of each species $j$. Dividing the overlaps of $i$ and $j$ by the total tiles of $j$, which we denote as $T_j$, will remove this bias. We define the overlap density of species $i$ with $j$ as

\begin{equation}
    \theta_{ij} \equiv \frac{O_{ij}}{T_j}
    \label{eq:thetaij}
\end{equation}
\noindent
Then we can compute the average overlap density.

\begin{equation}
    \overline{\theta_i} = \frac{\sum_{j=0}^{N_i}\theta_{ij}}{N_i}
    \label{eq:thetaoverlinei}
\end{equation}
\noindent
If the null hypothesis is correct, we expect the overlap density for species $i$ to be the average overlap over all the other species that we are comparing species~$i$ with. Therefore

\begin{equation}
    \mu_i/T_i =  \overline{\theta_i} 
    \label{eq:muioverti}
\end{equation}
\noindent
Therefore

\begin{equation}
    \mu_i = T_i\frac{\sum_{j=0}^{N_i} ( \frac{O_{ij}}{T_j})}{N_i}
    \label{eq:main_mui}
\end{equation}
\noindent
if the data for species $i$ was drawn from the same spatial Poisson process.

To assess whether the observed overlap $O_{ij}$ is to be expected typically, we use a $z$-score. To get the $z$-score we need to get the standard deviation $\sigma$. Assuming the data is drawn from a spatial Poisson process, the variance $\sigma^2$ equals the mean $\mu_i$. Therefore the $z$-score goes as follows

\begin{equation}
    z = \frac{O_{ij}-\mu_i}{\sqrt{\mu_i}}
    \label{eq:zscore}
\end{equation}
\noindent
This measures how far $O_{ij}$ is from its expected value $\mu_i$ in standard-deviation units. If $z$ is sufficiently large, we reject the null hypothesis.

Given this equation to ascertain the $z$ value, we only need to compute $\mu_i$ and $O_{ij}$. But in making the comparison between species $i$ and $j$, we must choose which of the species is $i$ and which is $j$, as the mathematical meaning of $z$ is altered if we exchange $i$ and $j$. In order to choose one we make a reasonable assumption that the rarer species inhabiting less tiles has a greater chance of relying on the more common species with more tiles. Therefore we choose the $i$ species as the species with less tiles, $T_i < T_j$, as it should yield more accurate results.

In addition to a $z$-score which indicates significance, a level of correlation can also be quantified. In this case it is the overlap over the mean \cite{Veech2013}.

\begin{equation}
    \rho_{ij} = \frac{O_{ij}}{\mu_i}
    \label{eq:rhocorr}
\end{equation}

Fine tuning may be needed depending on the dataset. The subsequent results are based on a dataset where I removed species that inhabited 5 or less tiles from the dataset. The reason being that a species with such few data points will not be able to yield useful results and it may very well skew the means in an unfavorable way. Another adjustment made was omitting avian observations from the dataset. This was done because birds tend to have less obligate relations with other species. For avian species, this method is less likely to yield as many interesting interactions as are found with invertebrate species.

\section{Results} 

This method was run on a dataset centered around Santa Cruz California (see~Fig.~\ref{fig:map_yellow_purple_SC}). It was bound by the latitudes $37.2551$ and $36.9209$ and the longitudes $-121.464$ and $-122.446$. The study area contains $4,684,582$ tiles, arranged $3313\times1414$. This corresponds to \SI{109.329}{km} (east–west) by \SI{46.662}{km} (north–south).

This dataset was downloaded in June of 2025 from iNaturalist\cite{inat_export_2025}. It includes all recorded non-avian species and genus-level taxa from the given area that inhabit at least six or more tiles. Observations with obscured or private locations were omitted from this dataset as well.

The data analysis on this dataset was performed on an 8-GB MacBook Air (M1). The quality and insight gained should increase with larger data sets and this will be explored in future work.

\subsection{Radial distribution function}

A quantitative way of understanding the correlations between different species is to use the radial distribution function (RDF) \cite{burgot2016radial}. The RDF for two species, $i$ and $j$, gives the probability per unit area that for any observation of $i$, an observation of $j$ is at a radius $r$ from it. To compute the RDF over a maximum distance of $R$, one determines all distances between observations for $i$ and observations for $j$ and bins them according to this distance, normalizing by the total number of pairs of distance less than R and the  area spanned by the corresponding bin. This gives the probability per pair, per unit area.

\begin{figure}[tbh]
	\centering
	\includegraphics[width=\linewidth]{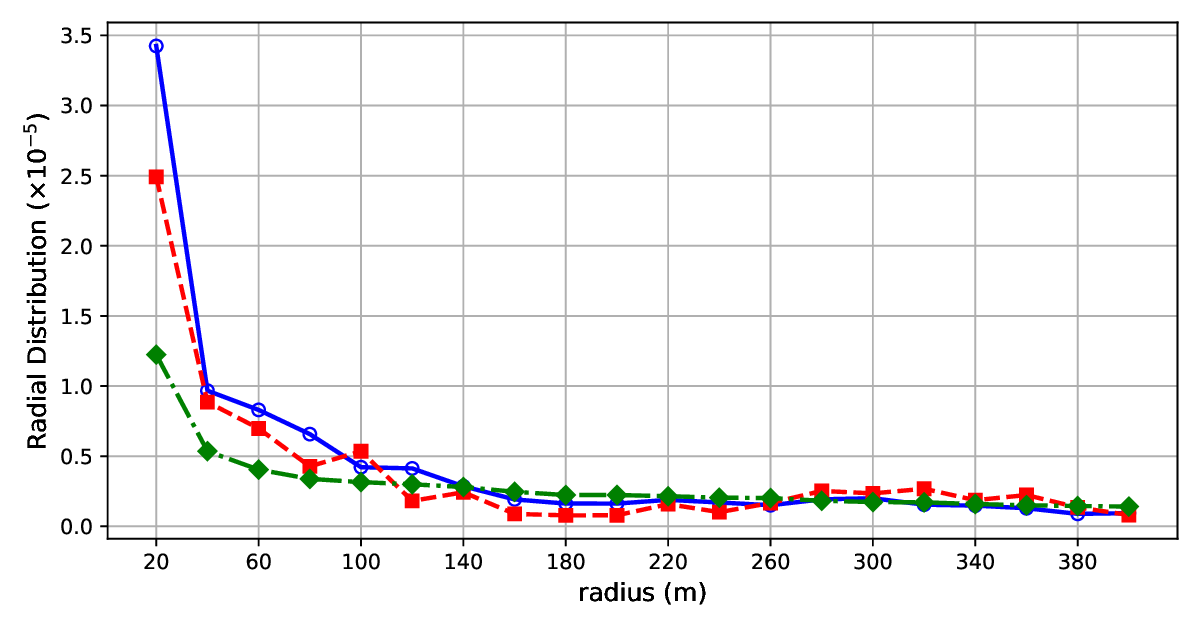}
	\caption{Normalized Radial Distribution functions for three species pairs.\\
    Blue: California Yerba Santa $\times$ Yerba Santa Beetle\\
    Red: Pacific Poison Oak $\times$ Yerba Santa Beetle\\
    Green: Coast Redwood $\times$ Slender Banana Slug}
	\label{fig:rad_dist}
\end{figure}

Here I have highlighted three different species pairs to illustrate typical and atypical correlations. One of the strongest associations, that will be further discussed below, is between California Yerba Santa with the Yerba Santa Beetle, (solid line) in Fig. \ref{fig:rad_dist}. The other two pairs, Pacific Poison Oak with Yerba Santa Beetle (dashed line) and Coast Redwood with Slender Banana Slug (dot dash) have weaker correlations that are still quite substantial. These distributions are typical of what one expects for species correlations and do not imply any  biological association. The heightened probability density seen for small radii are due to the aforementioned factors, that species observations occur along trails, paths, and in more biologically favorable areas.

\subsection{Tables}

\begin{table}[tb]
  \centering
  \begin{threeparttable}
    \caption{The top eleven known inter-species correlations within our dataset sorted by significance value (Sig). The results are displayed in order of highest significance descending. The percentile column indicates the rank within the entire output dataset.}
    \label{tab:known-relations}  
    \rowcolors{1}{rowgray}{white}
    \begin{tabularx}{\linewidth}{L c r r r}
      \toprule
        \rowcolor{white}
        Species ($i$) & Overlapping Species ($j$) & Sig. ($z$) & Corr. ($\rho$) & (\%) \\
        \midrule
        \showrowcolors
        \textit{Tamalia glaucensis} & Big Berry Manzanita & 37.52 & 21.24 & 99.98 \\
        Coyote Brush Bud Gall Midge & Coyote Brush & 30.42 & 9.18 & 99.96 \\
        Coffeeberry Midrib Gall Moth & Coffeeberry & 29.92 & 26.84 & 99.92 \\
        Coyote Brush Stem Gall Moth & Coyote Brush & 28.94 & 12.26 & 99.90 \\
        Sagebrush Woolly Stem Gall Midge & California Sagebrush & 28.64 & 13.65 & 99.88 \\
        White Sage Leaf Gall Midge & Black Sage & 28.28 & 44.07 & 99.86 \\
        Ceanothus Bud Gall Midge & Wartleaf Ceanothus & 28.28 & 44.07 & 99.84 \\
        Pumpkin Gall Wasp & Coast Live Oak & 25.25 & 8.98 & 99.75 \\
        Red Cone Gall Wasp & Valley Oak & 24.66 & 7.5 & 99.71 \\
        Yerba Santa Beetle & California Yerba Santa & 24.62 & 21.5 & 99.69 \\
        California Gall Wasp & Valley Oak & 24.31 & 5.46 & 99.65 \\ \bottomrule
    \end{tabularx}

  \end{threeparttable}
\end{table}

Table \ref{tab:known-relations} lists the top eleven known inter-species correlations within our dataset sorted by significance value, $z$. The rows are ordered from highest significance descending. It lists species $i$, species $j$, significance $z$, correlation $\rho$, and percentile ranking $\%$. Significance, correlation, and percentile are rounded to two decimal places.

The majority of the highest scoring significances are known obligate relations. In fact all but one of these from Table \ref{tab:known-relations} are cecidogenous (gall producing) species. This means whenever these are observed it is unlikely for them to be found in a spot without the host plant.

The Yerba Santa Beetle stands out for being the one species that is not cecidogenous. It represents many non-cecidogenous insect species that fully rely on a single host species / genus as a part of their life cycle. The Yerba Santa Beetle, \textit{Trirhabda eriodictyonis}, relies fully on species from the genus \textit{Eriodictyon} such as \textit{Eriodictyon californicum}, California Yerba Santa. It feeds exclusively on \textit{Eriodictyon} leaves as a larva and as an adult \cite{GouldWilson2015_EcolEvol}.

\begin{table}[tb]
  \centering
  \begin{threeparttable}
  \caption{The top six inter-species correlations between two species that share the same host plant.  The results are displayed in order of highest significance descending. The percentile column indicates the rank within the entire output dataset.}
  \label{tab:shared-host}
  \rowcolors{1}{rowgray}{white}
  \begin{tabularx}{\linewidth}{L c r r r}
    \toprule
      \rowcolor{white}
      Species ($i$) & Overlapping Species ($j$) & Sig. ($z$) & Corr. ($\rho$) & (\%) \\
      \midrule
      \showrowcolors
      Convoluted Gall Wasp & Red Cone Gall Wasp & 30.22 & 15.36 & 99.94 \\
      Spined Turban Gall Wasp & Red Cone Gall Wasp & 27.69 & 9.19 & 99.82 \\
      Club Gall Wasp & Disc Gall Wasp & 26.56 & 17.61 & 99.80 \\
      Yellow Wig Gall Wasp & Red Cone Gall Wasp & 26.05 & 13.83 & 99.78 \\
      Honeydew Gall Wasp & Red Cone Gall Wasp & 25.38 & 13.63 & 99.77 \\ \bottomrule
  \end{tabularx}

  \end{threeparttable}
\end{table}

Table \ref{tab:shared-host} lists the top six inter-species correlations between two species that share the same \textit{Quercus} host plant within our dataset sorted by significance value, $z$. The rows are ordered from highest significance descending. It lists species $i$, species $j$, significance $z$, and correlation $\rho$, and percentile $\%$. Significance, correlation, and percentile are rounded to two decimal places.

Some species end up correlating with other species because they have the same host plant. This was mainly seen with various oak associated gall wasps from the Tribe Cynipini. While strong user bias could be a reason for these results, there is a possibility that certain individual oaks have a greater chance of being infected by multiple cecidogenous wasp species \cite{Perea2021_GallIncidenceOaks}.

\begin{table}[tb]
  \centering
  \begin{threeparttable}
  \caption{The top three inter-species correlations between two species that do not appear to have any known inter-species interactions documented. The results are displayed in order of highest significance descending. The percentile column indicates the rank within the entire output dataset.}
  \label{tab:excess}
  \rowcolors{1}{rowgray}{white}
  \begin{tabular}{l l r r r}
    \toprule
      \rowcolor{white}
      Species ($i$) & Overlapping Species ($j$) & Sig. ($z$) & Corr. ($\rho$) & (\%) \\
      \midrule
      \showrowcolors
      Oregon Gumplant & Common Yarrow & 24.81 & 9.13 & 99.73 \\
      \textit{Strigamia} & Pacific Newts & 24.55 & 10.77 & 99.67 \\
      Rockweed & Ochre Sea Star & 24.16 & 6.8 & 99.63 \\ \bottomrule
  \end{tabular}
  
  \end{threeparttable}
\end{table}

Table \ref{tab:excess} lists the top three inter-species correlations within our dataset between two species that do not appear to have any known inter-species interactions documented, sorted by significance value, $z$. The rows are ordered from highest significance descending. It lists species $i$, species $j$, significance $z$, and correlation $\rho$. Both significance and correlation are rounded to two decimal places.

With complex ecosystems and observational biases in the data, the reason for some of these strong correlations is sometimes difficult to interpret. Because the coast is such a thin span of area, the coastal data from this dataset is too small to yield useful results. Therefore coastal relations should be ignored from these results. In the greater Santa Cruz area, Oregon Gumplant and Common Yarrow both generally inhabit coastal scrub ecosystems. It is likely the Rockweed and Ochre Sea Star association exists for a similar reason. If coastal ecosystems were to be explored more with this method, a much larger dataset would be needed.

\textit{Strigamia}, also found in Table \ref{tab:excess}, is a genus of Soil Centipedes (Order Geophilomorpha) that can be found in damp areas. It is not clear why it would correlate so strongly with Pacific Newts. One likely explanation is that in the Santa Cruz Mountains, next to the Lexington Reservoir, there is the Alma Bridge Road Newt Passage Project which has been documenting Pacific Newts and Pacific Newt roadkill instances along Alma Bridge Road \cite{midpen_newt_passage}. Throughout this \SI{5.4}{km} section of road, numerous unidentifiable Pacific Newt roadkill instances have been documented and marked as "Pacific Newts" on iNaturalist. Within this same section many \textit{Strigamia} roadkill have been observed as well. Because of the large scale of this citizen science project, the quantity of roadkill has caused a significant amount of co-occurrence for these species. It seems that because of this data anomaly, a strong association between \textit{Strigamia} and Pacific Newts has shown up in the results. A larger dataset should eliminate this artifact, unless there is a true unknown relation, which is possible but unlikely.

\section{Discussion}

This method is different from previous methods in the way it is able to filter correlations that are an artifact of the data. There are two kinds of reasons for why things appear to correlate when they do not. The first reason is user bias \cite{Geurts2023_SpatialBiasesCommunityScience}. The second reason is that species will be more prevalent in certain regions rather than others, but this does not mean to say that they interact \cite{BlanchetCazellesGravel2020_CooccurrenceNotEvidence, GobernaVerdu2022_CautionaryNotesCooccurrence}. In both cases a baseline is needed for what one would typically expect. Other work has not properly addressed this problem and will therefore yield inter-species interactions that are not as reliable. It is for this reason that this method is an important addition to modeling ecological data.

Additions to this method to characterize microhabitats for each species would be interesting. While a species of beetle might not correlate very highly with any single plant species, there is a chance it may find a particular combination of plants most habitable. That information could be revealed by the data given the right methodology. Incorporating time, elevation, and weather into a larger method could help yield further insight into ecosystem functionality.

There is also the potential to use this method to help in the delineation of ecoregions alongside algorithms that involve data clustering \cite{McInnes2017}.

\subsection{Exploring different estimates for the mean}
\label{sec:exploring_mui}

I have considered multiple ways of obtaining an average, $\mu_i$. Because this average is a crucial part of the methodology, it is important to explain why seemingly simpler alternatives fail. We can take a species $i$ and average every $O_{ij}$ for each overlapping species $j$. Let $N_i$ denote the number of $j$ species we compare to species $i$. The formula for the average overlap $\mu^{(O)}_i$ is

\begin{equation}
    \mu^{(O)}_i = \frac{\sum_{j=0}^{N_i}O_{ij}}{N_i}
    \label{eq:badavg}
\end{equation}

The reason for not using this average is because of the very broad distribution of species abundance that is best described by a log series distribution \cite{Baldridge2016_SADComparisonPeerJ}. The degree to which a species' tiles overlap will correlate strongly with their abundance. The very broad form of this abundance distribution will make the average of the mean abundance much larger than other measures, such as the median. This means that species with high abundance will dominate the mean in Eq. \ref{eq:badavg} because of high value outliers. This leads to a greatly reduced mean statistical power compared to the alternative definition Eq. \ref{eq:main_mui} that mitigates these outlier effects. In that formula the averaging of $O_{ij}$ employed a division by $T_i$ inside the computation of the average, which suppresses the effects of these high abundance outliers.

\section{Conclusion}
This paper has introduced a highly efficient method of getting likely candidates for strong inter-species interactions that can be extracted from citizen science datasets. This was done by constructing a method for filtering out specious correlations and devising a computational method that is very efficient. This method was run on a dataset of $15,975$ species, and $874,263$ data points, which has potentially $127,592,325$ interactions, using only modest computer resources. The results obtained find known interactions and some other ones that are likely artifactual but may actually have some ecological significance.

Future work could include: looking at spatial-temporal correlations, more comprehensive analysis of the entire database, using similar methods to better delineate biomes, and the analysis of species microhabitat.

\section{Acknowledgments}

I thank Prof. James Helfield and Prof. Merrill Peterson for useful discussions, and J.M. Deutsch for useful discussions and a critical reading of the manuscript.

\section*{Code availability}
All analysis code is available at \url{https://github.com/Jacob-Deutsch-Work/Finding-Inter-species-Associations-on-Large-Citizen-Science-Datasets}

\bibliographystyle{unsrtnat} 
\bibliography{references}

\end{document}